\documentclass[%
 aip,
 amsmath,amssymb,
 reprint,%
]{revtex4-1}

\usepackage{graphicx}
\usepackage{dcolumn}
\usepackage{bm}

\usepackage[utf8]{inputenc}
\usepackage[T1]{fontenc}
\usepackage{mathptmx}
\usepackage{etoolbox}
\usepackage{color}
\usepackage{hyperref}
\usepackage{tabularx}
\usepackage{makecell}
\usepackage[figuresleft]{rotating}
\usepackage{xcolor}
\usepackage{soul}

\include{vpack}

\makeatletter
\def\@email#1#2{%
 \endgroup
 \patchcmd{\titleblock@produce}
  {\frontmatter@RRAPformat}
  {\frontmatter@RRAPformat{\produce@RRAP{*#1\href{mailto:#2}{#2}}}\frontmatter@RRAPformat}
  {}{}
}%
\makeatother
\begin{document}

\preprint{AIP/123-QED}

\title{Importance of equivariant and invariant symmetries for fluid flow modeling}

\author{Varun Shankar}
\affiliation{Carnegie Mellon University}

\author{Shivam Barwey}
\affiliation{Argonne National Laboratory}

\author{Zico Kolter}
\affiliation{Carnegie Mellon University}

\author{Romit Maulik}
\affiliation{Argonne National Laboratory}
\email{rmaulik@anl.gov} 

\author{Venkatasubramanian Viswanathan}
\affiliation{Carnegie Mellon University}
\email{venkvis@cmu.edu}

\date{\today}

\begin{abstract}
Graph neural networks (GNNs) have shown promise in learning unstructured mesh-based simulations of physical systems, including fluid dynamics. In tandem, geometric deep learning principles have informed the development of equivariant architectures respecting underlying physical symmetries. However, the effect of rotational equivariance in modeling fluids remains unclear. We build a multi-scale equivariant GNN to forecast fluid flow and study the effect of modeling invariant and non-invariant representations of the flow state. We evaluate the model performance of several equivariant and non-equivariant architectures on predicting the evolution of two fluid flows, flow around a cylinder and buoyancy-driven shear flow, to understand the effect of equivariance and invariance on data-driven modeling approaches. Our results show that modeling invariant quantities produces more accurate long-term predictions and that these invariant quantities may be learned from the velocity field using a data-driven encoder.
\end{abstract}


\maketitle

\section{Introduction}
Numerical solutions have enabled insights into many real-world fluid phenomena \cite{pope}.  However, turbulent fluid flow remains particularly challenging to solve, and thus these methods are handicapped by their computational cost despite enormous progress in computational hardware over the last few decades\cite{Moin1998,Cant2002}. Novel deep learning methods offer the promise of breaking this scaling trade-off of computational cost and accuracy.\citep{Lu2021,neuraloperator,chen2018} 
Computational fluid dynamics (CFD) methods stand to gain significantly from AI-enhanced algorithms as more complex scientific and engineering flows test the limits of current numerical approaches \citep{Brunton2020}. Data-driven modeling has deep roots in CFD, demonstrated, for example, by reduced order modeling techniques such as proper orthogonal decomposition \cite{Deane1991,Cazemier1998,Rowley2004,Weiss_2019}, 
but machine learning (ML) has provided new opportunities for model design, and many popular deep learning approaches
have been explored for fluids applications in recent years. The success of convolutional neural network (CNN) architectures in computer vision has led to an unsurprising parallel effort in convolutional-type architectures for spatiotemporal scientific problems. Graph neural networks (GNNs), which may be considered an extension of CNNs to arbitrary grids, are particularly well-suited to fluids problems, where data are often represented on a computational mesh with arbitrary structure. Several works have studied the efficacy of graph representation learning for fluids applications, with encouraging results \cite{pmlr-v119-sanchez-gonzalez20a,Chen2021,NEURIPS2022_Allen,klimesch2022simulating}.

These machine learning models do not respect  underlying physical symmetries, e.g. rotation, and fail relatively simple equivariance tests where a rotated input does not produce an appropriately rotated output.  Geometric deep learning offers a pathway to invoke relevant group symmetries to the machine learning architecture, and through Noether's theorem \cite{Baados2016}, conservation laws can be mapped directly to symmetries, enabling the possibility of symmetry-respecting conservation-law obeying machine learning models.  

Fluid systems exhibit Euclidean symmetry, and in this work we are concerned with the use of spatial and rotationally equivariant graph neural network models. We examine four model architectures to understand the critical effect of embedded equivariance on modeling invariant and tensor-valued representations of fluid data. We assess model performance with regards to both accuracy and computational cost to provide a full picture of modeling strategies. We find that invariant representations of the flow state are effective for long-term forecasts of the flow field. Absent existing invariant representations, a neural network can be used to encode the invariant representation.  The core contributions of our work are: (i) We build a rotation and translation equivariant graph neural network for forecasting spatiotemporal fluid flow and compare with non-equivariant model baselines. (ii) We demonstrate that modeling rotation invariant as opposed to non-invariant flow features results in better prediction accuracy for both equivariant and non-equivariant networks. (iii) We show that embedded equivariance allows for more accurate forecasts and that encoding a latent state of invariant features is a viable method to significantly reduce the cost of equivariant architectures without sacrificing accuracy.

\section{Background}
\subsection{Graph neural networks}

Graph neural networks have gained traction in the deep learning community for applications where the data contain some underlying graph or network structure. GNNs can be viewed as a more general form of convolutional neural networks, which require the use of structured grid data. Instead of restricting convolutional kernels to operate on a specific choice of data structure, e.g. pixels in an image, GNNs expand the modeling range of deep learning architectures to data with arbitrary structure. Graphs are described via a series of nodes and edges, or connections between two nodes. Nodes may describe any data point in a group -- an atom in a molecule, a social network participant, or the value of a field at a point in space. Edges connect these nodes and allow for some measure of locality or influence that one node may have on another. By employing convolution or message-passing operations through aggregation of node features along edges, information can traverse the graph to generate deep graph features. Given the flexibility of this data structure, GNNs have been applied to a variety of regression and classification modeling tasks, including object detection \cite{Shi_2020_CVPR}, traffic modeling \cite{Bui2021}, drug discovery \cite{Jiang2021}, and recommendation algorithms \cite{Wu2019}.

Recently, more attention has been turned towards predicting PDE solutions with GNNs due to their discretization-independent capabilities \cite{NEURIPS2020_multipole,Shukla2022,li2020neural,iakovlev2021learning}. By representing an input field with an arbitrary point cloud or mesh, one may learn the solution operator using a deep graph network. This approach is predicated on the assumption that the solution field at a point is dependent on not only the localized input, but also on inputs within a neighborhood of influence at that point. GNN layers have deep connections to our traditional numerical methods for PDEs that validate this assumption. Finite volume techniques, for example, compute solutions by integrating fluxes over cell boundaries, a type of message-passing algorithm that draws parallels to GNN formulations. GNNs can also be linked to integral transforms \cite{goswami2022physicsinformed}, a convenient tool that is often used to solve complex differential equations.
An integral transform of the form:
\begin{equation}
    (Tu)(\vec{x}) = \int_{D_k} k(u(\vec{x}),u(\vec{s}),\vec{x},\vec{s})u(\vec{s})d\vec{s},
\end{equation}
where $\vec{x}, \vec{s} \in \mathbb{R}^d$ and $D_k$ is the domain of integration, can be discretized via Monte-Carlo integration to produce
\begin{equation}
    (Tu)(\vec{x}_i) \approx \frac{1}{N}\sum_{j=1}^N k(u(\vec{x}_i),u(\vec{x}_j),\vec{x}_i,\vec{x}_j)u(\vec{x}_j),
\end{equation}
which is analogous to message-passing schemes used in GNNs.

Graph-based architectures have therefore found a natural application in data-driven modeling of fluid flows, where data are typically obtained from numerical PDE solutions on unstructured computational meshes. The grid-independent framework is essential for problems in complex domains where a structured grid representation would require interpolation or other transformations. Indeed, GNNs have been used in a variety of fluid modeling contexts. Graph representations have been leveraged for modeling Lagrangian dynamics \citep{Li2018,Ummenhofer2020Lagrangian,pmlr-v119-sanchez-gonzalez20a}, steady-state predictions have been tackled in various works \citep{Yang2022}, including with solvers-in-the-loop \citep{belbute-peres20a}, and attention-style mechanisms have shown improvements in turbulence modeling \citep{Peng2022}. \citet{meshgraphnets} lays the foundation for unsteady PDE forecasting on graphs with derivative efforts \citep{lino2022} employing similar Encode-Process-Decode architectures.

\subsection{Equivariance}

Equivariant neural networks are increasingly finding use in scientific machine learning applications. Many of the physical systems we are interested in modeling respect certain symmetries, and it can be convenient to encode these symmetries directly into the network. A model or function map that is equivariant with respect to a certain symmetry group commutes with the action of the symmetry group, i.e. applying a symmetry transformation to the function output is equivalent to applying the transformation to the function input and then evaluating the function. If the target function being modeled is equivariant, it holds that the model should be equivariant as well. How best to achieve this model equivariance, however, is open to investigation. A common approach in machine learning is data augmentation. Applying symmetry transformations to the data and labels during training allows a model to implicitly learn the symmetry relationships withing the data, but ultimately equivariance cannot be guaranteed. Instead, a more invasive approach is to specifically design equivariant architectures that require no data augmentation. The hope is that by hard-constraining models to obey appropriate symmetries, we can promote better generalization and data efficiency, which has been validated in a variety of modeling contexts \cite{wang2020,Batzner2022,pmlr-v162-mondal22a,Gong2022,Burby2020}.

The vast number of symmetry groups means that many common deep learning architectures are already equivariant with respect to certain symmetries. CNNs, for example, are equivariant to translations and GNNs are permutation equivariant. For physical systems in Euclidean space, rotation equivariance is another important symmetry that can be connected to conservation of angular momentum \cite{Baados2016}. Rotation equivariance has been less straightforward to implement in deep learning, but a number of advanced architectures such as E(2)-CNNs \cite{weiler2021general}, spherical CNNs \cite{esteves17_learn_so_equiv_repres_with_spher_cnns}, and equivariant GNNs \cite{pmlr-v139-satorras21a} have been put forth in recent years.
Rotation equivariant graph networks and tensor-field networks in particular have demonstrated value in accelerating molecular and N-body particle dynamics simulations \citep{tfn,Batzner2022}. However, exploration of equivariance's role and application of these neural architectures to fluids problems has remained sparse in literature. Nevertheless, some studies leveraging equivariance and invariances for modeling fluid problems have been conducted. \citet{ling_kurzawski_templeton_2016} and \citet{gao} use rotation invariant and equivariant networks for turbulence modeling. Rotation equivariant CNNs \citep{wang2020} and GNNs \citep{lino2022,suk2022} have also demonstrated improved predictive accuracy in forecasting tasks.

\section{Methods}
\label{sec:methods}
\begin{figure*}
\begin{center}
\includegraphics[width=0.8\textwidth]{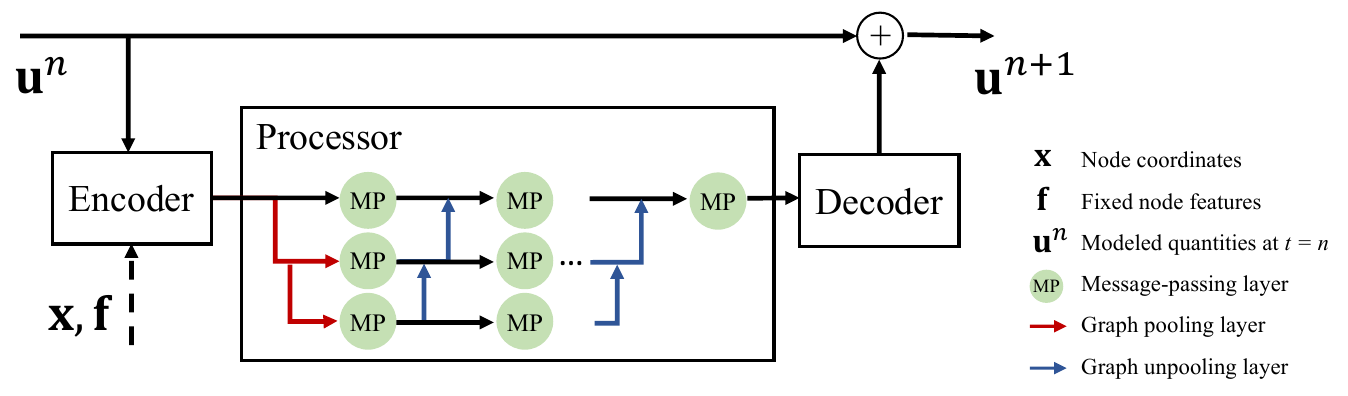}
\end{center}
\caption{Schematic of the overall model architecture, indicating encoder and decoder blocks enclosing the multi-level message-passing graph processor. Inputs to the network at each timestep are $\mathbf{u}^n$, the current dynamical quantities, $\mathbf{x}$, the node coordinates, and $\mathbf{f}$, non-modeled quantities such as boundary conditions and external forces.}
\label{fig:schem}
\end{figure*}

The model follows similar design principles as existing literature on learning simulations on graphs, such as \citet{meshgraphnets}. The system is described at a point in time by a point cloud given by coordinates $\mathbf{x}_i \in {\mathbf{x}_0,...,\mathbf{x}_N}$, fixed node features $\mathbf{f}_i$ corresponding to boundary conditions, external force fields, or global flow characteristics, e.g. Reynolds number, and modeled quantities $\mathbf{u}_i$, such as velocity and temperature. The model predicts the state of the system at the next time step $t+1$ and any future time steps $t+n$ using an iterative integrator. 
The model uses the Encode-Process-Decode paradigm, illustrated in Fig. \ref{fig:schem}, that has shown to be effective for modeling dynamical systems. The encoder generates the graph and transforms inputs into latent features. The latent features are processed with a series of message-passing layers and decoded to produce the update quantity passed to the integrator. The primary goal of the study is to investigate the effect of embedded equivariance on model performance. Thus, we evaluate several model architectures that follow this general framework, described subsequently.

\subsection{Equivariant operations}
In this section, we first introduce some basic concepts needed to build deep equivariant networks that operate on tensor-valued fields such as velocity. 
We recognize Euclidian rotations and translation as the set of symmetries to embed in our models. While there are other important symmetries in fluid motion, such as Galilean invariance, we do not consider them here. In principle, invariance and equivariance to symmetries is straightforward. Given a roto-translation operator $R$, an invariant network $f_\theta$ obeys the following:
\begin{equation}
    f_\theta(x) = f_\theta(Rx),
\end{equation}
and an equivariant network $f_\theta$ obeys the following:
\begin{equation}
    Rf_\theta(x) = f_\theta(Rx),
\end{equation}
where the parameters $\theta$ are unchanged on both sides of the equation. In other words, we can apply a transformation to the input or the output and receive equivalent results. By extension, any composition of equivariant functions is also equivariant:
\begin{align}
    Rf_\theta(x) &= f_\theta(Rx)\\
    Rg_\theta(x) &= g_\theta(Rx)\\
    Rg_\theta(f_\theta(x)) &= g_\theta(Rf_\theta(x)) = g_\theta(f_\theta(Rx)).
\end{align}
Thus, to construct a deep equivariant neural network, it is sufficient to ensure that each of the layer operations in the network are equivariant.

All of the equivariant operations within the network can be generalized to a form of the tensor product. The tensor product, represented as
\begin{equation}
    a \otimes b ,
\end{equation}
is a bilinear, equivariant map between two vector spaces. We illustrate the product with a relevant example. If $a$ and $b$ are three-dimensional vectors in Euclidian space, we can represent them with a Cartesian basis:
\begin{equation}
    \vec a = \sum_{i=1..3}a_i\vec x_i, \quad \vec b = \sum_{i=1..3}b_i\vec x_i.
\end{equation}
The tensor product is given by:
\begin{equation}
    \vec a \otimes \vec b= \sum_{i=1..3}\sum_{j=1..3}a_ib_j\vec x_i \otimes \vec x_j,
\end{equation}
where $\vec x_i \otimes \vec x_j$ forms the basis for the tensor product vector space. This tensor product has dimensionality $3\times3=9$ and can also be recognized as the outer or dyadic product of $a$ and $b$. Here, we have chosen a particular representation of the tensor by means of our basis, but we may also perform a change of basis to form another representation. 

The Cartesian tensor representation is a \textit{reducible} representation with respect to three-dimensional Euclidean group transformations. We discuss the notion of a group element, which can be expressed as a mapping between a group element $g$, such as a rotation around axis $\vec r$ by angle $\gamma$, to a matrix $D(g)$ that acts on a certain representation. A representation is reducible if a change in basis transforms $D(g)$ to block diagonal form, meaning the group element acts independently on nontrivial subrepresentations of the representation. In the case of a scalar, which is invariant to any rotations, the group element is given by the $1\times1$ matrix $[\mathbf{1}]$. It is straightforward to note that this is an \textit{irreducible} representation, or irrep. For a Cartesian vector, the group element is given by the familiar $3\times3$ rotation matrix $R(\vec r,\gamma)$. This basis also forms an irreducible representation. Moving to the nine-dimensional Cartesian tensor, the group element is a $9\times9$ matrix formed by the tensor product of rotation matrices $R(\vec r,\gamma) \otimes R(\vec r,\gamma)$. However, this representation is reducible because in contains group invariant linear subspaces. The trace of the tensor, also given by the dot product of the two vectors $\vec a \cdot \vec b$, is invariant to rotations and transforms as a scalar with the $1\times1$ group element. Similarly, the antisymmetric component of the tensor, also given by the cross product of the vectors $\vec a \times \vec b$, transforms as a vector with the standard $3\times3$ rotation matrix. Lastly, the symmetric, traceless component of the tensor transforms as a 2-rank tensor with a $5\times5$ group element. Thus, the $3\times3$ Cartesian tensor can be decomposed into the direct sum of these three irreducible representations -- the isotropic, deviatoric, and antisymmetric components of the tensor. With the appropriate change of basis, the group element for the whole representation becomes a block diagonal matrix with $1\times1$, $3\times3$, and $5\times5$ blocks corresponding to the group elements of the irreducible representations. We use irreps to represent data within the network, which simplifies computation of the equivariant operations and makes generalization to higher-order tensors and arbitrary representations straightforward.

The equivariant architecture is underpinned by three core equivariant operations: equivariant linear transformations, fully-connected tensor products, and gated non-linearities. 

The equivariant linear transformation linearly combines tensors of the same rank. The $i$-th rank-$l$ tensor output is given by
\begin{equation}
    z_i^l = \sum_j w_{ijl} x_j^l ,
\end{equation}
where we denote any tensor-valued quantities with the superscript $l$. The weights of the layer, $w_{ijl}$, are indexed by output channels $i$, input channels $j$, and tensor rank $l$. The weights are restricted to scalar values, but $x_i^l, z_i^l$ could represent scalars, vectors, or higher-order tensors. In principle, this is a familiar operation in deep learning models. The difference is that for non-scalar features $x_i^{l>0}$, we must multiply all components of a tensor-valued feature by the same scalar weight to preserve equivariance:
\begin{align*}
    Rz_i^l &= R(\sum_j w_{ijl} x_j^l) \\
    &= \sum_j R(w_{ijl} x_j^l) \\
    &= \sum_j w_{ijl} (Rx_j^l)
\end{align*}
One clear disadvantage of this operation is that it does not allow for mixing of information between features of differing ranks; all outputs are computed in parallel across ranks. This necessitates use of the more general tensor product.

The fully-connected tensor product computes the product of two sets of irreps to form a third, arbitrary set of irreps. As discussed previously, the tensor product of two irreps $a$ and $b$ can be decomposed into a series of independent, equivariant, bilinear operations whose direct sum corresponds to $a\otimes b$. The allowable operations, denoted \textit{paths}, are dependent on the input irreps. For example, three paths comprise the tensor product of two vectors to output a scalar, vector, and symmetric traceless rank-2 tensor. If the inputs are representations of multiple irreps, potentially of different ranks, we can take the pairwise product of each irrep in the inputs to form a large set of possible output irreps using the allowable paths. The output irreps can then be linearly combined (within irreps) to produce an arbitrary output representation. Each of the paths are multiplied by a scalar weight, which represents the trainable parameters of the layer. Because each of the independent weighted path operations are equivariant, we can use this product to mix information between irreps while maintaining equivariance. We denote the tensor product as
\begin{equation}
    z = x \otimes (\mathbf{W}) \; y ,
\end{equation}
where $x$ and $y$ are input sets of irreps, $z$ is the output representation, and $\mathbf{W}$ are the learnable path weights.

Care must be taken when employing non-linear activation functions as a naive implementation of pointwise non-linearities would operate on each tensor component individually, a non-equivariant transformation. However, since the norm is an invariant quantity, the non-linearity can be applied to each irrep norm. Here, we use gated non-linearities, where the previous network layer outputs additional scalars for each irrep in the representation. These scalars are passed through an activation function and used to gate the norm of the output.

\subsection{Message-passing layers}
Nonlinear graph message-passing layers comprise the bulk of the model, especially within the processor. Following general GNN schemes, an edge update is performed before aggregating edge features and applying a node update. Equivariance is achieved by ensuring that both updates are equivariant transformations. We use multi-layer perceptrons (MLPs) with one hidden layer. These MLPs may be made equivariant by replacing standard linear layers with equivariant linear layers and standard non-linear activations in the hidden layer with equivariant gated non-linearities. Equivariant MLPs are denoted with $\text{MLP}_{eq}$.

Given incoming edge features $\mathbf{h}_{ij}$ and node features $\mathbf{h}_{i}$, first the edge update is computed:
\begin{align}
    \mathbf{v}_{ij} &= f_v(\mathbf{h}_{ij},\mathbf{h}_{i},\mathbf{h}_{j}) \\
    \mathbf{W} &= f_w(\mathbf{x}_{i},\mathbf{x}_{j}) \\
    \mathbf{h}_{ij}' &= \mathbf{v}_{ij} \otimes (\mathbf{W}) \; Y_{l\leq 2}(\mathbf{r}_{ij}) , \label{eq:tp}
\end{align}
where $f_v,f_w$ are trainable functions, $\otimes (\mathbf{W})$ denotes a fully connected tensor product with weights $\mathbf{W}$, $\mathbf{r}_{ij}$ is the relative position vector $\mathbf{x}_{i}-\mathbf{x}_{j}$, and $Y(\mathbf{r}_{ij})$ are spherical harmonics, up to $l_{max}=2$. The weights of the tensor product are given as a function of the node coordinates $\mathbf{x}_{i},\mathbf{x}_{j}$, but the exact definition of the kernel function $f_w$ varies with model class. If the input and output representations are equivalent, the edge update is a residual update:
\begin{equation}
    \mathbf{h}_{ij} \leftarrow \mathbf{h}_{ij} + \mathbf{h}_{ij}' ,
\end{equation}
otherwise, if not equivalent:
\begin{equation}
    \mathbf{h}_{ij} \leftarrow \mathbf{h}_{ij}' .
\end{equation}

\begin{figure}
\centering
\includegraphics[width=0.48\textwidth]{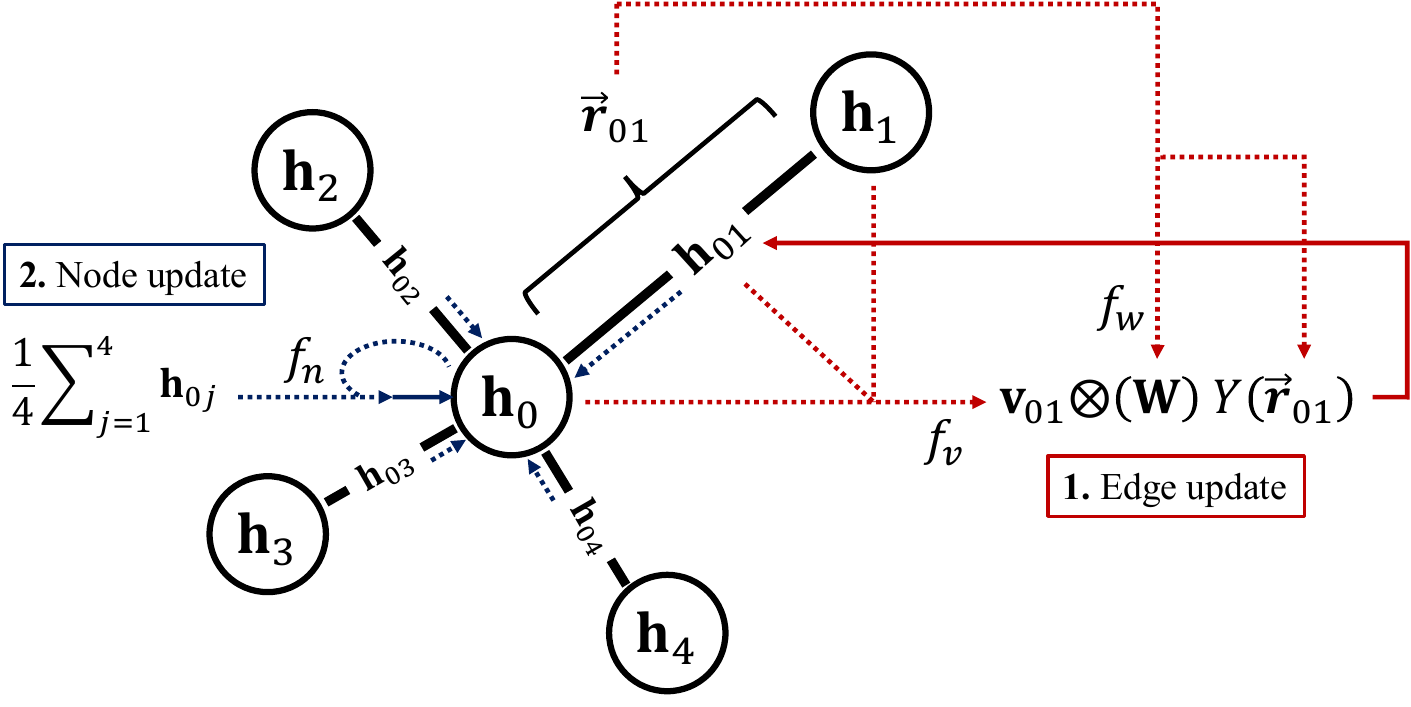}
\caption{Illustration of an example message-passing step. The edge update is computed from current node and edge features and includes the fully-connected tensor product. The node features are updated using the aggregated edge features and current node features.}
\label{fig:mp}
\end{figure}

The new edge features are used to update the node features. The edge features are aggregated via summation over a neighborhood of connectivity $\mathcal{N}$ and used as input to another network:
\begin{align}
    \mathbf{h}_{i}' &= f_n(\frac{1}{|\mathcal{N}|}\sum_\mathcal{N}\mathbf{h}_{ij},\mathbf{h}_{i}).
\end{align}
Again, a residual update is used if input and output representations are equivalent. We illustrate the general message passing algorithm in Fig. \ref{fig:mp}.

Three different forms of the message-passing algorithm are used in our experiments. $f_v, f_w$, and $f_n$ are varied, along with implementation details for computational efficiency. For the equivariant message-passing layer, $f_v$ and $f_n$ are $\text{MLP}_{eq}$'s and $f_w$ is a function of the form
\begin{equation}
    f_w = \text{MLP}(\text{Embed}(||\mathbf{x}_{i}-\mathbf{x}_{j}||)), 
    \label{eq:fw_iso}
\end{equation}
where the edge length $||\mathbf{x}_{i}-\mathbf{x}_{j}||$ is embedded via projection onto a 16-dimensional set of basis functions. Since the edge length is invariant to rotation, it is sufficient to use an MLP as opposed to $\text{MLP}_{eq}$.

In the non-equivariant formulation, there is no distinction made between scalar features and vector or higher-order tensor features; each tensor component is treated as an independent scalar. As a result, eq. \ref{eq:tp} can be simplified to a linear transformation
\begin{equation}
    \mathbf{h}_{ij}' = \mathbf{v}_{ij} \mathbf{W}^T , 
\end{equation}
equivalent to the fully-connected tensor product between two sets of scalars to within a proportional factor. We use this expression directly to avoid tensor product computational overhead. $f_v$ and $f_n$ are MLPs instead of $\text{MLP}_{eq}$'s, and we adjust the $f_w$ MLP inputs to account for the loss of directional information from $Y(\mathbf{r}_{ij})$,
\begin{equation}
    f_w = \text{MLP}(\text{Embed}(||\mathbf{r}_{ij}||), \mathbf{x}_{i},\mathbf{x}_{j}), 
    \label{eq:fw_aniso}
\end{equation}
providing both the embedded edge length and the node coordinates directly.

A key differentiating factor in the non-equivariant layer is that eq. \ref{eq:fw_aniso} enables anisotropic tensor product weights. If we remove the $\mathbf{x}_{i},\mathbf{x}_{j}$ inputs as in eq. \ref{eq:fw_iso}, the weights are isotropic, which is necessary to develop an equivariant layer. Thus, a third layer, the isotropic layer, can be constructed, where $f_v$ and $f_n$ are again MLPs and $f_w$ is given by eq. \ref{eq:fw_iso}. If all features consist of $l=0$ scalar irreps, this layer is in fact also equivariant, while still leveraging the optimized compute kernels for traditional linear or dense layers. While the equivariant layer is generalizable to any feature representation, the isotropic layer requires exclusively scalar representations everywhere to ensure a rotationally equivariant operation.

We tabulate the key differences between message-passing layers in tab. \ref{tab:mp}.
\def\arraystretch{1.3}
\begin{table}[]
\caption{Comparison of message-passing layers.}
\begin{tabular}{rlll}
\hline
\multicolumn{1}{l}{} & $f_v$             & $f_w$ inputs                                         & $f_n$             \\ \hline
\textbf{Equivariant}          & $\text{MLP}_{eq}$ & $||\mathbf{r}_{ij}||$                                & $\text{MLP}_{eq}$ \\
\textbf{Non-equivariant}      & MLP               & $||\mathbf{r}_{ij}||, \mathbf{x}_{i},\mathbf{x}_{j}$ & MLP               \\
\textbf{Isotropic}            & MLP               & $||\mathbf{r}_{ij}||$                                & MLP               \\ \hline
\end{tabular}
\label{tab:mp}
\end{table}

\subsection{Encoder/Decoder}
The encoder transforms all inputs into both latent edge and latent node features for processing. Only nodal information $\mathbf{x}_i,\mathbf{f}_i,\mathbf{u}_i$ is provided as input. Thus, first a graph must be generated from the node coordinates $\mathbf{x}_i$. When learning from simulations on a computational mesh, often the mesh connectivity is directly used to generate the graph representation. However, we take a more general approach to edge generation by using a radial cutoff. This removes some of the implicit bias of the discretization, particularly for meshes with high aspect ratio or skewed cells, and more importantly, does not require that the data come from a computational mesh, suitable for example, for learning from experimental data. The radial cutoff is an additional hyperparameter that allows for more control over the sparsity of the graph's adjacency matrix. Since only a point cloud of data is required, it is also possible to subsample the computational mesh to decrease data size.

\begin{figure}
\centering
\includegraphics[width=0.45\textwidth]{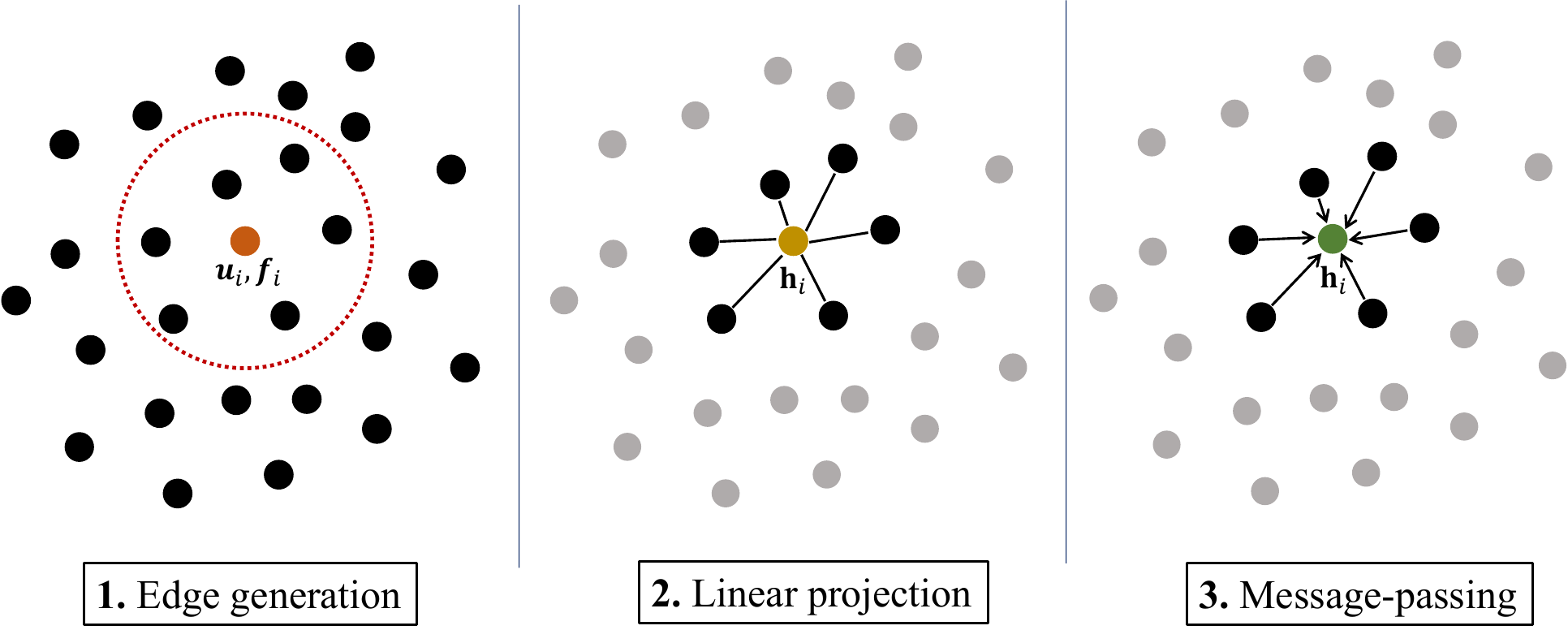}
\caption{Visualization of the encoding procedure. Edges are generated from a radial cutoff using the node coordinates. Node and edge inputs are linearly projected to a high-dimensional hidden state. One message-passing layer computes the resulting latent state of arbitrary representation.}
\label{fig:enc}
\end{figure}

Thus, first edges with relative position vector $\mathbf{r}_{ij}$ are generated using a radial cutoff. Two steps complete the transformation -- a node- and edge-wise linear projection to the hidden layer and one message-passing layer to the output latent space. Initial node features are given by $\mathbf{f}_i,\mathbf{u}_i$, while initial edge features are the spherical harmonics $Y(\mathbf{r}_{ij})$, up to $l_{max}=1$. Fig. \ref{fig:enc} depicts this encoder process. In the equivariant case, the linear projection is the equivariant linear layer and the message-passing layer is the equivariant message-passing layer, otherwise, a standard linear layer and the non-equivariant message-passing layer are used for the non-equivariant encoder. The invariant encoder is almost identical to the equivariant encoder, except the outputs are only scalars. We tabulate the differences in encoders in Tab. \ref{tab:enc}. It is common for the encoder to be restricted to pointwise projections, however, our focus on the latent state representation requires a more complex encoder algorithm. To learn "differential-type" operators for generating the latent representation, we include the message-passing layer to incorporate spatial information. For example, the vorticity field is computed from the curl of the velocity field, which cannot be evaluated from a pointwise operation if only the velocity field is provided.

The decoder mirrors the encoder architecture, however, only the node features are decoded. One message-passing layer and a linear projection realize the transformation. Again, the decoder can also be made equivariant or non-equivariant by selecting the appropriate linear and message-passing layers. 
The output is passed to the integrator to produce the system state at the next time step.

\begin{table}[]
\caption{Comparison of encoder modules.}
\begin{tabular}{rlll}
\hline
\multicolumn{1}{l}{} & Projection & Message-passing & Outputs\\ \hline
\textbf{Equivariant}          & Eq-linear  & Equivariant     & Scalars/vectors\\
\textbf{Invariant}          & Eq-linear  & Equivariant     & Scalars\\
\textbf{Non-equivariant}     & Linear     & Non-equivariant & N/A\\ \hline
\end{tabular}
\label{tab:enc}
\end{table}

\subsection{Graph pooling}
Research efforts on graph representation learning have motivated the development of various graph coarsening or pooling strategies \citep{rethinkpooling2020,pmlr-v196-chen22a,graphUnets}. When data are represented on a physical grid in space, pooling can capture multi-scale features that have been shown to be effective for better convergence and accuracy of models. This formalism is used extensively in CNN-based architectures, particularly U-net architectures \cite{ronneberger2015u}, which have seen widespread use in scientific machine learning applications. The U-net approach generates increasingly coarser representations of the data from convolutional pooling operations, capturing and evolving features at different length-scales. During the unpooling step, fine-grained features are concatenated back to the data to generate new features that capture multiscale information. The benefit of this approach is that long-range interactions can be represented without composing many small-scale convolutions, which add significant computational complexity. 

The challenge with graph pooling is that unlike uniform grids, there is no well-defined natural hierarchical method to coarsen the data structure. As such, this leaves many potential pooling strategies open for study. The goal of the pooling step is to generate a new, coarser graph representation on which to perform additional message-passing. Approaches used in literature can generally be grouped into methods that purely target the spatial structure of the graph and methods that leverage the learned graph features to produce a new or pruned graph. Among these spatial methods is the use of voxel-clustering algorithms \cite{lino2022}, which overlay a voxel grid over the domain, pooling nodes within each voxel to generate a coarse graph node at the centroid of the voxel. Here, the coarse graph length scale is well-defined from the size of the voxel. Feature-based approaches rely on graph node or edge attributes instead of coordinate values and are often used for graph data that do not correspond to a spatial domain, e.g. social networks. Top-$k$ pooling \cite{meshgraphnets} is one such method that computes a score for each node, sorting the nodes and pruning node indices greater than $k$, resulting in a smaller graph that learns to efficiently represent the larger high-level graph.

\begin{figure}
\centering
\includegraphics[width=0.45\textwidth]{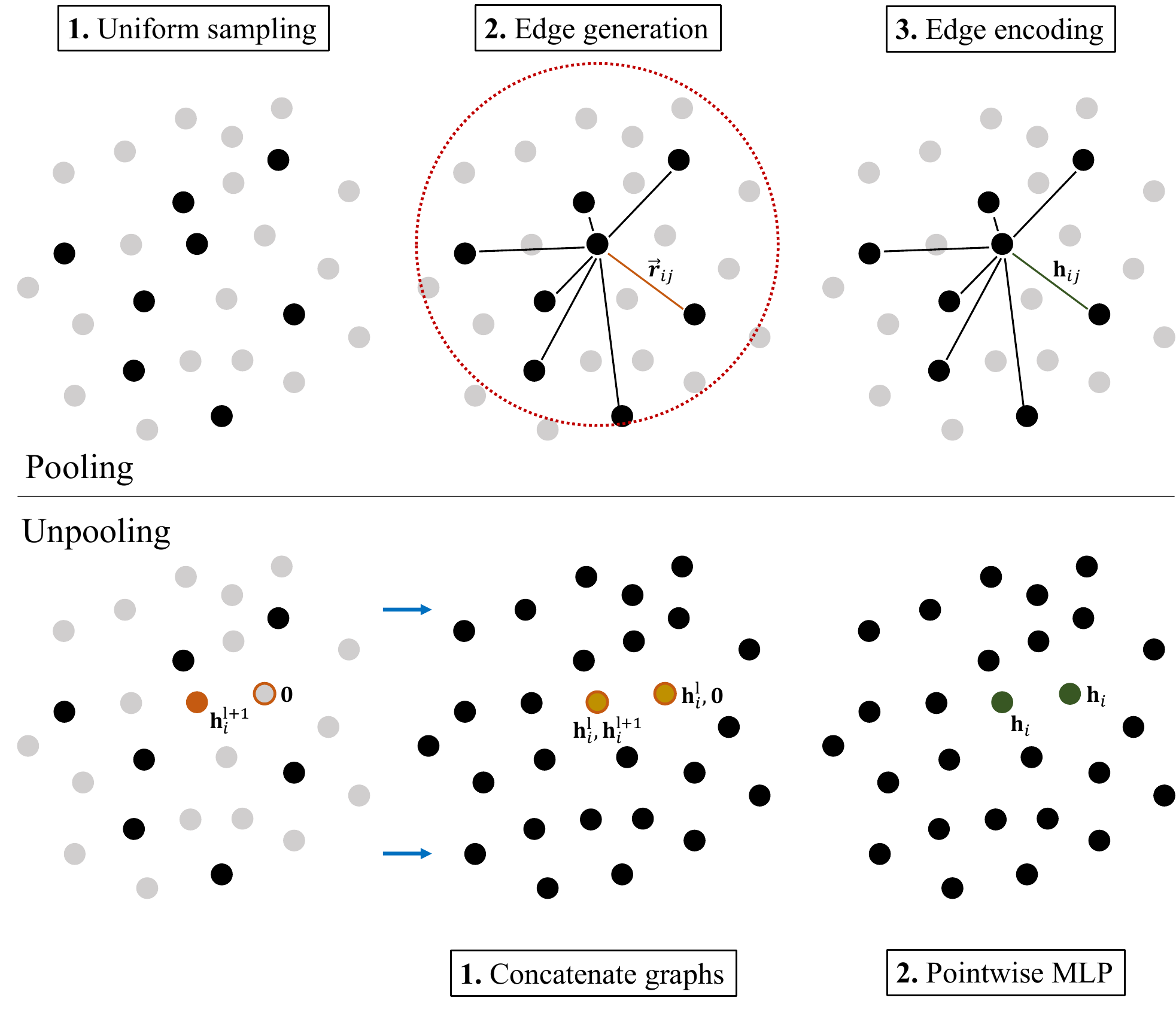}
\caption{Illustration of the graph pooling and unpooling layers. The pooling layer samples the top-level point cloud, generates edges from a radial cutoff, and encodes the new edges. The unpooling layer populates an empty feature matrix corresponding to the top-level nodes with lower-level features and concatenates them with the top-level features. A pointwise MLP reduces the dimensionality back to the latent space.}
\label{fig:pool}
\end{figure}

Our approach is remarkably simple and efficient, requiring zero knowledge of the graph structure or attributes. A random subgraph is generated using uniform sampling of the top-level graph nodes. We can capture larger length scale interactions because the sampled subgraph is sparser than the top-level graph while still being proportional to the original point cloud distribution. While the node features are carried over to the coarse graph, new edges are generated from a larger radial cutoff and latent edge features encoded from spherical harmonics $Y(\mathbf{r}_{ij})$ of the relative position vectors with an MLP (or $\text{MLP}_{eq}$). The unpooling operation distributes lower-level node features to an empty feature matrix on the top-level graph, while a node-wise MLP (or $\text{MLP}_{eq}$) computes new top-level features as $\text{MLP}(\mathbf{h}_i^l,\mathbf{h}_i^{l+1})$, where $l$ and $l+1$ denote the top-level and lower-level graphs respectively. This can be viewed as a delta-function interpolation, with subsequent top-level message-passing layers taking on the task of further distributing lower-level information. Fig. \ref{fig:pool} illustrates the pooling and unpooling algorithms.

\subsection{Models}
The purpose of this work is to investigate the role equivariance and invariance plays in modeling fluid flows with GNNs. A major aspect we explore is the consequence of embedded equivariance on model performance. Since the underlying fluid datasets are generated from physics models that respect Euclidean symmetries, we expect embedded equivariance to improve the prognostic capabilities of the model. However, a significant impediment to more widespread adoption of equivariant GNN architectures is the increased computational burden these models impose. This is particularly salient in tensor field networks and similar architectures that leverage tensor-valued data representations, as equivariance is embedded through the use of tensor products, which are substantially more computationally complex than the linear operations common in many deep learning architectures. A further consequence of sparse adoption is that tensor product compute kernels have not been fully refined and optimized to take advantage of GPU hardware acceleration, in contrast to e.g. Dense or CNN layers, which have received much attention over the past decade. As a result, researchers must find a balance between compute cost and model performance, weighing the benefits of equivariance with the increased computational cost it contributes.

Our model design aims to combat this issue by developing an additional equivariant architecture that imposes only a minor increase in computational cost relative to non-equivariant GNNs. Since most of the cost originates with the tensor product, we seek to avoid the use of tensor products as much as possible. Tensor products are only required if the data or latent representations involve tensor-valued quantities. Since data inputs in fluid modeling often contain vector or higher order data, either from the velocity field or external forces such as gravity, we focus our attention on encoding an invariant scalar latent representation, such that the processor need only consist of isotropic message-passing layers, thus avoiding tensor products in the majority of the evaluation. The motivation for using invariant quantities in fluid modeling and analysis is already well-established. The turbulent kinetic energy, turbulent dissipation rate, and enstrophy are a few examples of important scalar fluid dynamical quantities that can be used to effectively describe a fluid system. By learning a high-dimensional scalar latent representation of the flow state, we can still model the flow dynamics using an equivariant model with limited computational overhead.

Various combinations of the encoder/decoder and processor blocks are combined to produce three distinct model classes. The first equivariant model, denoted as \textbf{eq}, encodes latent features comprising scalar and vector-valued quantities using the equivariant encoder/decoder and equivariant processor. This takes the tensor field network \cite{tfn} approach to modeling and thus adds considerable cost to training and evaluation. It has been demonstrated that latent representations with higher-order tensors can improve predictive capacity even further \cite{Batzner2022}, however, we limit ourselves to $l_{max}=1$ as training models with $l_{max}>1$ was computationally infeasible. The second, alternate equivariant model (\textbf{eqscl}) generates scalar invariant latent features using the equivariant encoder, but the processor consists of isotropic message-passing layers, which provides a computational advantage. Here, we can avoid tensor products in the processor and model the dynamics of the learned invariant latent representation. We additionally test two non-equivariant models. The non-equivariant model (\textbf{neq}) trades all equivariant blocks for non-equivariant blocks. Thus, no equivariance is imposed in the model design. This architecture is duplicated for the second model (\textbf{neqaug}), but equivariance is included by augmenting the training data with random rotations. The models are compared in Tab. \ref{tab:mod}.

\begin{table}[]
\caption{Comparison of model types.}
\begin{tabular}{rlll}
\hline
  & Encoder & Processor MP   & Data aug. \\ \hline
\textbf{neq}    & Non-equivariant & Non-equivariant & False             \\
\textbf{neqaug} & Non-equivariant & Non-equivariant & True              \\
\textbf{eq}     & Equivariant     & Equivariant     & False             \\
\textbf{eqscl}  & Invariant     & Isotropic       & False             \\ \hline
\end{tabular}
\label{tab:mod}
\end{table}

\subsection{Training}
The models are fit to the datasets by minimizing the loss function,
\begin{equation}
    \mathcal{L}_\theta(\mathbf{u}(\mathbf{x}_i, t)) = \frac{1}{N}\sum_{i=0}^N (\mathbf{u}(\mathbf{x}_i, t+\Delta t)-\widetilde{\mathbf{u}}_\theta(\mathbf{x}_i, t+\Delta t))^2 ,
\end{equation}
where $\mathbf{u}(\mathbf{x}_i, t)$ are the ground truth modeled quantities evaluated at $N$ nodal coordinates $\mathbf{x}_i$ and time $t$, and $\widetilde{\mathbf{u}}_\theta(\mathbf{x}_i, t+\Delta t)$ are the predicted quantities from a model parameterized by $\theta$ at coordinates $\mathbf{x}_i$ and time $t+\Delta t$, where $\Delta t$ is the forecast time step. We use a one-step rollout during training, meaning the model only predicts the fluid field at one $\Delta t$ beyond the input field, equating to one function evaluation. After the model is trained, we evaluate the generalization accuracy on much longer predictions using an iterative rollout. Longer rollouts during training have been shown to improve accuracy of long-horizon predictions, however, the training cost scales with the size of the rollout. Instead, one may approximate the effects of a long prediction by corrupting the input fields with noise. Model predictions will inevitably contain some error, which can accumulate and lead to instabilities during iterative rollouts. By teaching the model to be robust to noise, we can enable more stable long horizon predictions. In our experiments, we use random noise with mean 0 and variance 0.001, which is added to the input field $\mathbf{u}(\mathbf{x}_i, t)$ at each training iteration. 

Models are implemented in PyTorch using the \texttt{e3nn} package \citep{e3nn} for equivariant operations and \texttt{PyG} \citep{Fey/Lenssen/2019} for graph-specific functionality. Models are trained with the Adam optimizer using a decaying learning rate varying from $10^{-3}$ to $10^{-5}$ on 8 V100 GPUs. In addition, stochastic weight averaging is used in the final 10\% of training, where model weights are averaged over the last 10\% of epochs to promote better generalization.

\section{Experiments}
We demonstrate the model performance on predicting the spatiotemporal evolution of two examples of canonical two-dimensional fluid flows. The first example targets the discretization-independent capabilities of the model, using unstructured mesh data from a finite-volume simulation of incompressible flow around a cylinder. We see that the GNN can adapt to arbitrary grids including random point clouds obtained from sampling the computational mesh. In the second example, we examine more complex multiphysics flow phenomena described by the Boussinesq equations. Although we fix the discretization to the original uniform grid from simulation, we still observe benefits of the equivariance-embedded architecture.

We compute three metrics to evaluate model accuracy. We report the average one-step mean-squared-error (MSE) over samples in the test set, the same loss that is used for training. No input noise is used in the evaluation phase. Additionally, we perform an iterative rollout prediction over the entirety of the test trajectory and report the coefficient of determination ($R^2$), given by
\begin{equation}
    R^2 = 1-\frac{\sum_{j}\sum_{i} (\mathbf{u}(\mathbf{x}_i, t_j)-\widetilde{\mathbf{u}}_\theta(\mathbf{x}_i, t_j))^2}{\sum_{j}\sum_{i} (\mathbf{u}(\mathbf{x}_i, t_j)-\bar{\mathbf{u}}(\mathbf{x}_i, t_j))^2} ,
\end{equation}
where $\bar{\mathbf{u}}$ is the mean of the ground truth dataset and the squared errors are summed over all spatial locations $\mathbf{x}_i$ and time steps $t_j$.
Finally we evaluate the equivariance loss, an unsupervised loss given by
\begin{equation}
    \text{MSE}(Rf_\theta(\mathbf{u}_0, \mathbf{x}, \mathbf{f}),f_\theta(R\mathbf{u}_0, R\mathbf{x}, R\mathbf{f})),
\end{equation}
where $f_\theta$ is the model, $R$ is a random rotation transformation, and $\mathbf{u}_0$ (initial condition), $\mathbf{x}$ (node coordinates), $\mathbf{f}$ (fixed node features) represent all inputs to the network.

\subsection{Cylinder Flow}
\begin{figure*}
  \begin{center}
      \includegraphics[width=\linewidth]{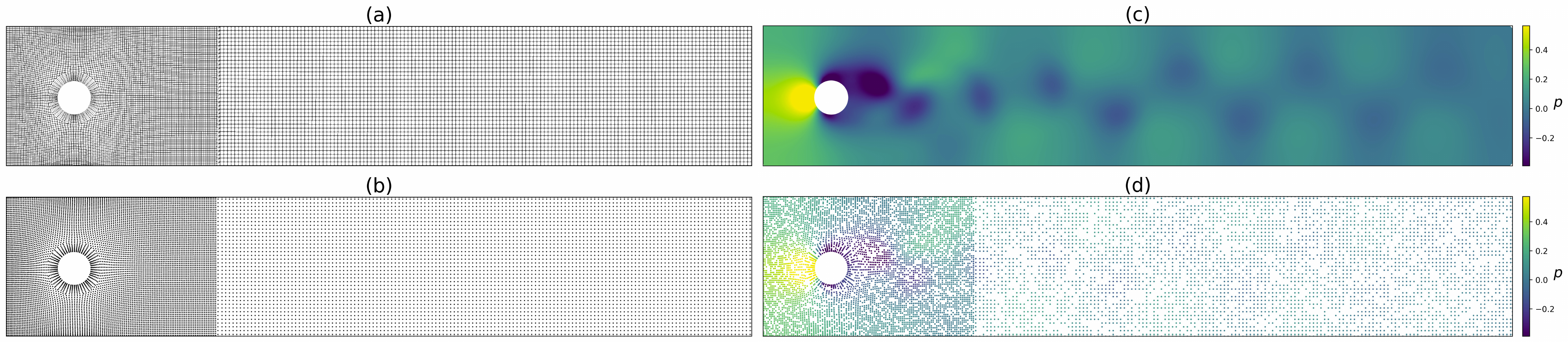}
  \end{center}
  \caption{(a) Computational mesh used for simulating the cylinder flow case. (b) Point cloud obtained from cell centers of the computational mesh. (c) Snapshot of the pressure field from numerical simulation. (d) Example of the sampled pressure point cloud provided to the model during training.}
  \label{fig:cyl_data}
\end{figure*}

The first dataset consists of flow around a cylinder confined in a channel at a low Reynolds number $Re=600$, following the incompressible Navier-Stokes equations:

\begin{align}
\frac{\partial \mathbf{u}}{\partial t}+(\mathbf{u}\cdot\nabla)\mathbf{u}&=\frac{1}{Re}\nabla^2\mathbf{u}-\nabla p \\ \nabla\cdot\mathbf{u}&=0.
\label{eq:ns-nd}
\end{align}

The computational domain contains a cylinder of diameter $D=0.1$ placed within a channel of height $H=0.41$ and length $L=2.2$. The top, bottom, and cylinder boundaries are no-slip walls and the right boundary is a pressure outlet. The left boundary is set to a velocity inlet, assuming a fully-developed laminar parabolic velocity profile with $U_{max}=1.5$. The computational mesh is depicted in Fig. \ref{fig:cyl_data}(a). The simulation was conducted using the open-source library \texttt{OpenFOAM} with pressure-corrector solver \texttt{pimpleFOAM} to enforce continuity and second-order numerical schemes. The simulation is run for 8 seconds, with the first 3 seconds discarded to allow for vortex-shedding to develop.

The flow is solved using the velocity-pressure formulation, but post-processing enables recovery of the vorticity and streamfunction fields as well. The fluid state can be represented by the pressure field and either the velocity field or the vorticity and streamfunction. The vorticity and streamfunction are scalar invariant fields in two-dimensions and thus do not transform under rotation, in contrast to the velocity field. We investigate the effect of modeling both representations in addition to model architecture choices. The data are coded with tags \textbf{wsp} or \textbf{uvp} for modeling vorticity, streamfunction, pressure or velocity and pressure respectively.

The point cloud used for generating the graph is obtained from the cell centers of the computational mesh. These nodes are illustrated in Fig. \ref{fig:cyl_data}(b). The model is provided an initial set of fields as input, such as the pressure field shown in Fig. \ref{fig:cyl_data}(c), along with fixed node features -- a one-hot vector encoding the boundary condition type. During training, the point cloud is additionally uniformly sampled to restrict the size of the training sample to between 8000 and 10000 nodes, as shown in Fig. \ref{fig:cyl_data}(d). The model is trained on a one-step-ahead prediction over three seconds of flow between $t=3$ and $t=6$. The test set consists of flow between $t=6$ and $t=8$. Two model hyperparameters are specific to each dataset, the number of message-passing layers in the processor and the dimensionality of the latent space. For the cylinder dataset, we choose 4 message-passing layers and a latent dimensionality of 64.

\subsection{Marsigli Flow}
We also evaluate our approach on a canonical strong-shear flow that exhibits the Kelvin-Helmholtz instability, known as Marsigli flow \citep{Ahmed2021}. Two fluids of different densities, or equivalently temperatures, are contained in a channel separated by a barrier. Once the barrier is removed, the fluids mix according to the Boussinesq equations. The Boussinesq equations leverage the assumption that the fluid density differences are negligible, except when multiplied by gravity. Therefore, buoyancy-effects can still be represented while reducing the complexity of the governing equations. The 2D incompressible Boussinesq equations can be written in terms of the vorticity and streamfunction,
\begin{align}
\frac{\partial \omega}{\partial t}+(\mathbf{u}\cdot\nabla)\omega &= \frac{1}{Re}\nabla^2\omega+Ri\frac{\partial \theta}{\partial x} \\ 
\frac{\partial \theta}{\partial t}+(\mathbf{u}\cdot\nabla)\theta &= \frac{1}{RePr}\nabla^2\theta \\
\nabla^2\psi&=-\omega,
\label{eq:bouss}
\end{align}
where $\omega,\psi,\theta$ are the vorticity, streamfunction, and temperature respectively, and the dimensionless quantities that define the flow conditions are Reynolds number $Re$, Richardson number $Ri$, and Prandtl number $Pr$. The vorticity can be obtained from the velocity field as $\omega\mathbf{k}=\nabla \times \mathbf{u}$.

Marsigli flow is often used as an idealized case study for understanding ocean current dynamics. This is a particularly challenging problem for data-driven models due to the highly transient nature of the flow.
The computational domain is a box of length $L=8$ and height $H=1$ discretized on a uniform Cartesian grid of size $512 \times 64$ with free-slip boundary conditions. Numerical ground truth solutions are obtained using standard second-order central finite difference schemes. Again, we investigate the effect of modeling both vorticity-streamfunction and velocity fields. The data are coded with tags \textbf{wst} or \textbf{uvt} for modeling vorticity, streamfunction, temperature or velocity and temperature respectively.

The point cloud used for forecasts is downsampled from the original computational mesh by a factor of two in both directions. No sampling of the nodes is done during training. The model is provided an initial set of fields and fixed node features as before. In this case, the fixed node features were a one-hot encoding of the boundary conditions and, importantly, a vector representing the direction of gravitational forces, which was necessary to encode the anisotropy present in the system. The training dataset contains flow at four different $Re$, 700, 900, 1100, and 1300 for 8 seconds. The test set consists of flow at $Re=1000$. The complexity of the Marsigli system resulted in a need for a larger model compared to the cylinder flow. Here, we increase the number of processor message-passing layers to 6 and quadruple the latent dimensionality to 256.

\section{Results and discussion}

\begin{figure*}
  \begin{center}
      \includegraphics[width=\linewidth]{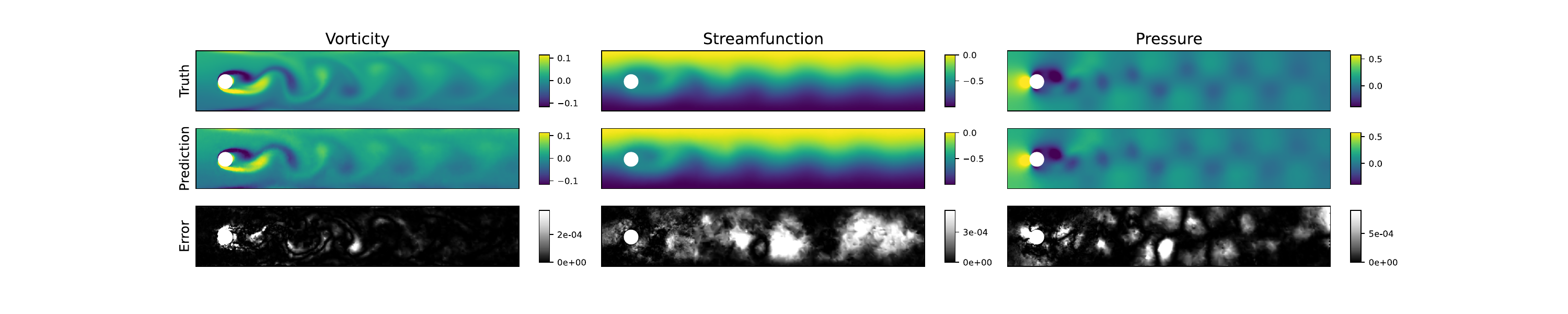}
  \end{center}
  \caption{Vorticity, streamfunction, and pressure fields of the cylinder flow dataset at the last timestep of the simulation, $t=8$. The ground truth is compared with predictions from equivariant model \textbf{wst-eqscl}.}
  \label{fig:cyl}
\end{figure*}

\def\arraystretch{1.2}
\begin{table}
\centering
\caption{Comparison of model error on the cylinder flow test set.}
\begin{tabular}{|p{0.8cm}|p{1.1cm}||p{1.5cm}|p{1.5cm}|p{2cm}|} \hline
\textbf{Data type} & \textbf{Model type} & \textbf{1-step test MSE} & \textbf{Forecast} $R^2$ & \textbf{Equivariance MSE} \\ \hline \hline
\textbf{wsp} & \textbf{neq}     & 2.73e-5 & 0.9758 & 4.72e-3 \\ \hline 
\textbf{wsp} & \textbf{neqaug}  & 3.94e-5 & 0.9863 & 2.96e-6 \\ \hline 
\textbf{wsp} & \textbf{eq}      & 3.10e-5 & 0.9968 & 3.21e-7 \\ \hline 
\textbf{wsp} & \textbf{eqscl}   & 3.23e-5 & 0.9925 & 3.92e-7 \\ \hline \hline 

\textbf{uvp} & \textbf{neq}      & 3.54e-5 & 0.9626 & 5.82e-3 \\ \hline 
\textbf{uvp} & \textbf{neqaug}   & 1.00e-4 & 0.8953 & 1.81e-5 \\ \hline 
\textbf{uvp} & \textbf{eq}       & 3.02e-5 & 0.9915 & 5.16e-7 \\ \hline 
\textbf{uvp} & \textbf{eqscl}    & 4.29e-5 & 0.9919 & 4.91e-7 \\ \hline 

\end{tabular}
\label{tab:cyl}
\end{table}

\def\arraystretch{1.2}
\begin{table}
\centering
\caption{Model timings for the cylinder flow case.}
\begin{tabular}{|p{0.8cm}|p{1.1cm}||p{1.5cm}|p{1.6cm}|} \hline
\textbf{Data type} & \textbf{Model type} & \textbf{Train time (hrs)} & \textbf{Evaluation time (s)}\\ \hline \hline
\textbf{wsp} & \textbf{neq}     & 4.5 & 1.14 \\ \hline 
\textbf{wsp} & \textbf{neqaug}  & 6.2 & 1.16 \\ \hline 
\textbf{wsp} & \textbf{eq}      & 11.6 & 3.83 \\ \hline 
\textbf{wsp} & \textbf{eqscl}   & 5.0 & 1.23 \\ \hline \hline 

\textbf{uvp} & \textbf{neq}      & 4.5 & 1.12 \\ \hline 
\textbf{uvp} & \textbf{neqaug}   & 6.8 & 1.11 \\ \hline 
\textbf{uvp} & \textbf{eq}       & 12.2 & 4.27 \\ \hline 
\textbf{uvp} & \textbf{eqscl}    & 5.6 & 1.87 \\ \hline 

\end{tabular}
\label{tab:cyl_time}
\end{table}

We first examine the results from the cylinder dataset. Fig. \ref{fig:cyl} shows snapshots of the vorticity, streamfunction and pressure fields obtained from the forecasted model prediction at the last timestep in the data, $t=8$, of model \textbf{eqscl} trained on \textbf{wst} data representations. We observe strong visual fidelity of the model, with the predicted trajectory able to accurately reproduce the wake dynamics of the cylinder flow. The results are summarized quantitatively in Tab. \ref{tab:cyl}. We report the 1-step mean-squared error (MSE) on the test set, the coefficient of determination $R^2$ on the full trajectory forecast, the unsupervised equivariance error, as well as train and evaluation times of each model in Tab. \ref{tab:cyl_time}. Metrics are computed using the pressure field, which is consistent across flow state representations. The 1-step test error is comparable across all models and data with errors of order $O(10^{-5})$. The one outlier is \textbf{uvp-neqaug} with MSE $1.00\times 10^{-4}$. 

The forecast $R^2$ shows more significant variation between models and data and is not strongly correlated with the training (1-step) accuracy. We observe greater predictive accuracy when modeling scalar variables \textbf{wsp} over vector-scalar data \textbf{uvp}. This trend holds true for each model investigated. The difference is most significant with model \textbf{neqaug}, which must implicitly learn the equivariant relationships in the data through augmentation of the training set. The result is unsurprising when one considers that in the \textbf{wsp} formulation, only the nodal coordinates $\mathbf{x}_i$ are transformed by the augmentation process, whereas with \textbf{uvp}, the modeled quantities $\mathbf{u}_i$ are also transformed.

We find that models \textbf{eq} and \textbf{eqscl} universally outperform models \textbf{neq} and \textbf{neqaug} in forecast accuracy. The equivariance-embedded architectures are able to achieve $R^2>0.99$ unlike the non-equivariant counterparts. The value of equivariance is highlighted especially by the interesting result from \textbf{wsp-neqaug}. Despite \textbf{wsp-neq} and \textbf{wsp-neqaug} having the same architecture, the learned equivariance from \textbf{wsp-neqaug} promotes better forecast generalization even though the 1-step test loss is larger than \textbf{wsp-neq}. We surmise that this behavior is not observed with \textbf{uvp} because it is more challenging for \textbf{neqaug} to adapt to rotations of the velocity data. Indeed, we see that the equivariance error of \textbf{wsp-neqaug} is about an order of magnitude lower than \textbf{uvp-neqaug}.

We can also compare the performance of models \textbf{eq} and \textbf{eqscl}. Model \textbf{wsp-eq} has the best forecast accuracy, outperforming \textbf{wsp-eqscl}, but the reverse is true for \textbf{uvp} data, although the accuracies are comparable. The most significant benefit to model \textbf{eqscl} is its computational efficiency. \textbf{eqscl} is able to achieve comparable accuracy to \textbf{eq} at less than half the cost in both train and evaluation time. The training time is additionally lower than \textbf{neqaug}, which was trained for longer than \textbf{neq} to reach convergence, due to its larger effective dataset size.

\begin{figure}
  \begin{center}
      \includegraphics[width=.8\linewidth]{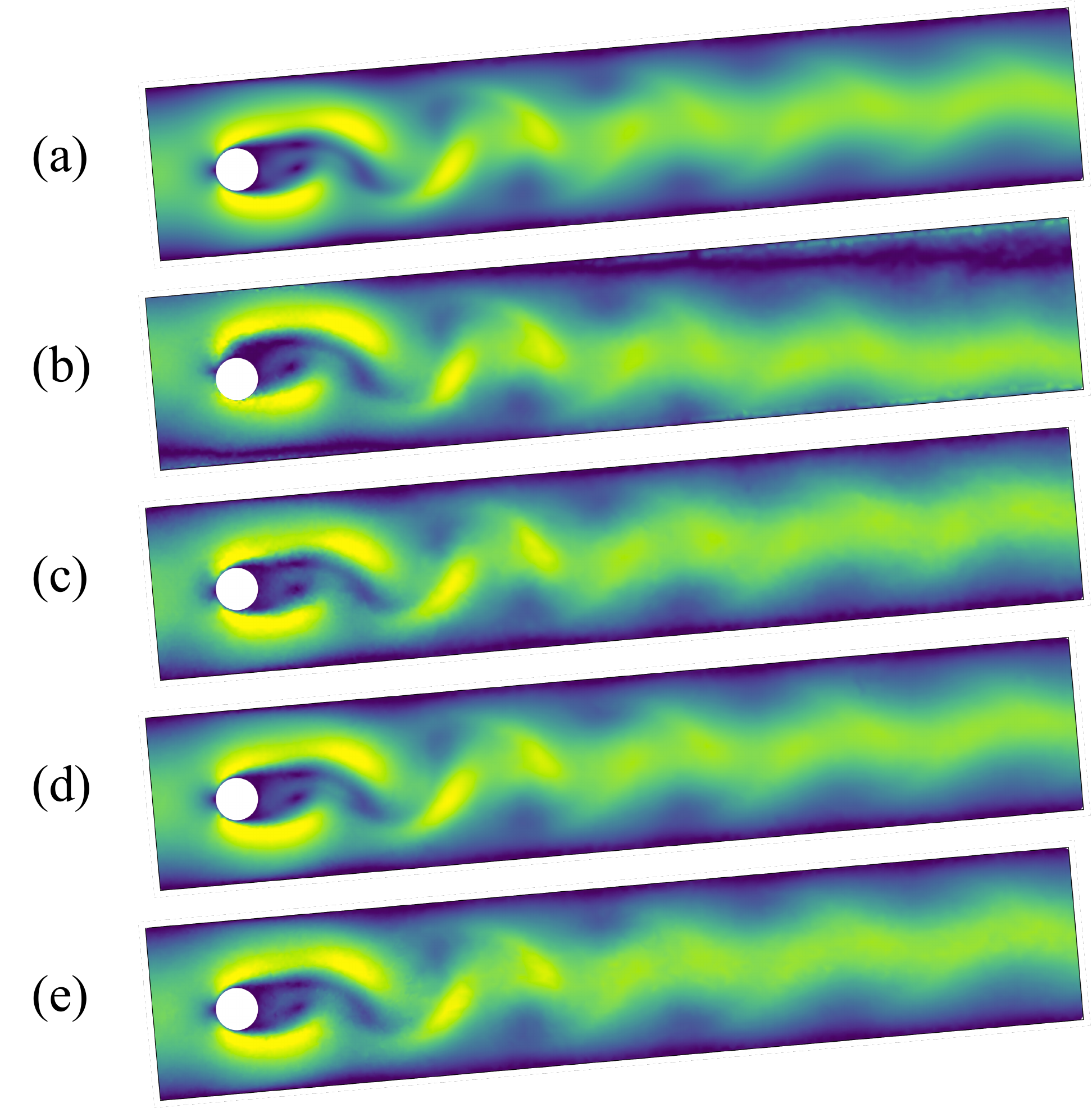}
  \end{center}
\caption{Velocity magnitude from a 5$^{\circ}$ rotation of the cylinder test set at $t=6.2$, obtained from (a) ground truth and predictions from models (b) \textbf{neq}, (c) \textbf{neqaug}, (d) \textbf{eq}, and (e) \textbf{eqscl}.}
  \label{fig:v_rot}
\end{figure}

The effect of rotation equivariance is illustrated in Fig. \ref{fig:v_rot}. We rotate the input data by 5 degrees counterclockwise and visualize the predicted velocity magnitude field at $t=6.2$ for the four \textbf{uvp} models. The two equivariant models and the data-augmented non-equivariant model are able to reproduce the rotated velocity field reasonably well in this qualitative comparison. However, model \textbf{neq} is clearly biased towards its training orientation, returning a horizontal wake and zero velocity at locations where the unrotated dataset contained a no-slip wall.

\begin{figure*}
  \begin{center}
      \includegraphics[width=\linewidth]{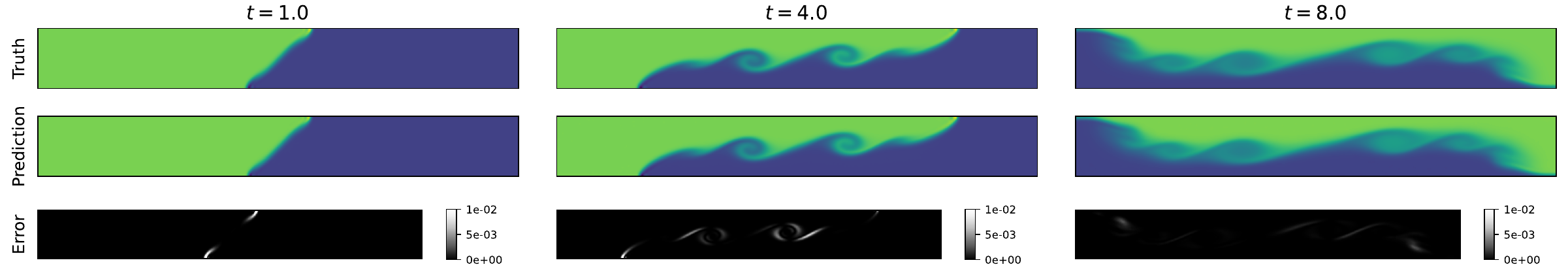}
  \end{center}
\caption{Snapshots of the temperature field from the Marsigli flow test case at various time steps. The ground truth is compared with predictions from equivariant model \textbf{wst-eqscl}.}
  \label{fig:msg}
\end{figure*}

\def\arraystretch{1.2}
\begin{table}
\centering
\caption{Comparison of model error on the Marsigli flow test set.}
\begin{tabular}{|p{0.8cm}|p{1.1cm}||p{1.5cm}|p{1.5cm}|p{2cm}|} \hline
\textbf{Data type} & \textbf{Model type} & \textbf{1-step test MSE} & \textbf{Forecast} $R^2$ & \textbf{Equivariance MSE} \\ \hline \hline
\textbf{wst} & \textbf{neq}     & 7.55e-7 & 0.9988 & 4.88e-3 \\ \hline
\textbf{wst} & \textbf{neqaug}  & 2.25e-6 & 0.9324 & 1.27e-6 \\ \hline
\textbf{wst} & \textbf{eq}      & 2.60e-6 & 0.9700 & 1.02e-7 \\ \hline
\textbf{wst} & \textbf{eqscl}   & 1.86e-6 & 0.9972 & 1.32e-7 \\ \hline \hline

\textbf{uvt} & \textbf{neq}      & 7.58e-7 & 0.9772 & 1.40e-3 \\ \hline
\textbf{uvt} & \textbf{neqaug}   & 2.57e-6 & 0.9487 & 8.41e-7 \\ \hline
\textbf{uvt} & \textbf{eq}       & 1.84e-6 & 0.9755 & 1.42e-7 \\ \hline
\textbf{uvt} & \textbf{eqscl}    & 7.75e-7 & 0.9863 & 9.20e-7 \\ \hline

\end{tabular}
\label{tab:msg}
\end{table}

\def\arraystretch{1.2}
\begin{table}
\centering
\caption{Model timings for the Marsigli flow case.}
\begin{tabular}{|p{0.8cm}|p{1.1cm}||p{1.5cm}|p{1.6cm}|} \hline
\textbf{Data type} & \textbf{Model type} & \textbf{Train time (hrs)} & \textbf{Evaluation time (s)}\\ \hline \hline
\textbf{wst} & \textbf{neq}     & 12.9 & 3.41 \\ \hline
\textbf{wst} & \textbf{neqaug}  & 19.5 & 3.40 \\ \hline
\textbf{wst} & \textbf{eq}      & 64.2 & 9.38 \\ \hline
\textbf{wst} & \textbf{eqscl}   & 19.3 & 3.94 \\ \hline \hline

\textbf{uvt} & \textbf{neq}      & 12.9 & 3.71 \\ \hline
\textbf{uvt} & \textbf{neqaug}   & 19.5 & 3.68 \\ \hline
\textbf{uvt} & \textbf{eq}       & 65.2 & 9.07 \\ \hline
\textbf{uvt} & \textbf{eqscl}    & 19.5 & 4.10 \\ \hline

\end{tabular}
\label{tab:msg}
\end{table}

For the Marsigli dataset, Fig. \ref{fig:msg} shows a qualitative comparison of temperature fields from the ground truth simulation and model \textbf{wst-eqscl}. We can observe the shear layer develop over time as the lower temperature, higher density fluid flows from the right under the lower density fluid to create the vortex sheet. Again, the model is able to reproduce the flow with visual fidelity, and as expected, the error is generally concentrated at the interface between the two fluids. In Tab. \ref{tab:msg}, we report the same metrics as before. Again, all models achieve comparable 1-step errors, however, differences are observed in the forecast $R^2$. Generally, modeling the invariant fields \textbf{wst} results in higher forecast accuracy, except for within the data-augmented models. This is a shift from the cylinder dataset, where \textbf{wst-nequag} significantly outperformed \textbf{uvp-nequag}. One of the reasons for this difference may be that in the Marsigli case, the fixed node features $\mathbf{f}_i$ are not just scalars, but also contain the vector representation of the direction of gravity. This adds another element to the equivariance that must be learned implicitly, even in the \textbf{wst} representation. Data-augmentation for this flow leads to a decrease in accuracy, similar to \textbf{uvp-nequag} in the cylinder case, although equivariance can be maintained.

Model \textbf{wst-neq} achieves the highest forecast accuracy of all models with \textbf{wst-eqscl} nearly matching its performance. For the \textbf{wst} data in this case then, the use of an equivariant inductive bias does not necessarily significantly impact long-term forecast accuracy, unless of course the generalization dataset includes rotated data, evidenced by the equivariance error for both \textbf{wst-neq} and \textbf{wst-eqscl}. \textbf{uvt-eqscl}, however, achieves better forecast accuracy than \textbf{uvt-neq}, indicating again that the use of vector data plays a role in equivariance's advantage.
Surprisingly, \textbf{eqscl} consistently outperforms \textbf{eq} in both cost and forecast accuracy, perhaps due to the invariant latent state representation instead of higher-order tensor representations. Since the model capacity of \textbf{eq} should be comparable to \textbf{eqscl}, this is a potential indicator that optimization of \textbf{eq} network parameters resulted in a local minimum instead of the global minimum. Given the extensive cost of training \textbf{eq}, it was infeasible to run experiments with more advanced optimization techniques.

\begin{figure}
  \begin{center}
      \includegraphics[width=\linewidth]{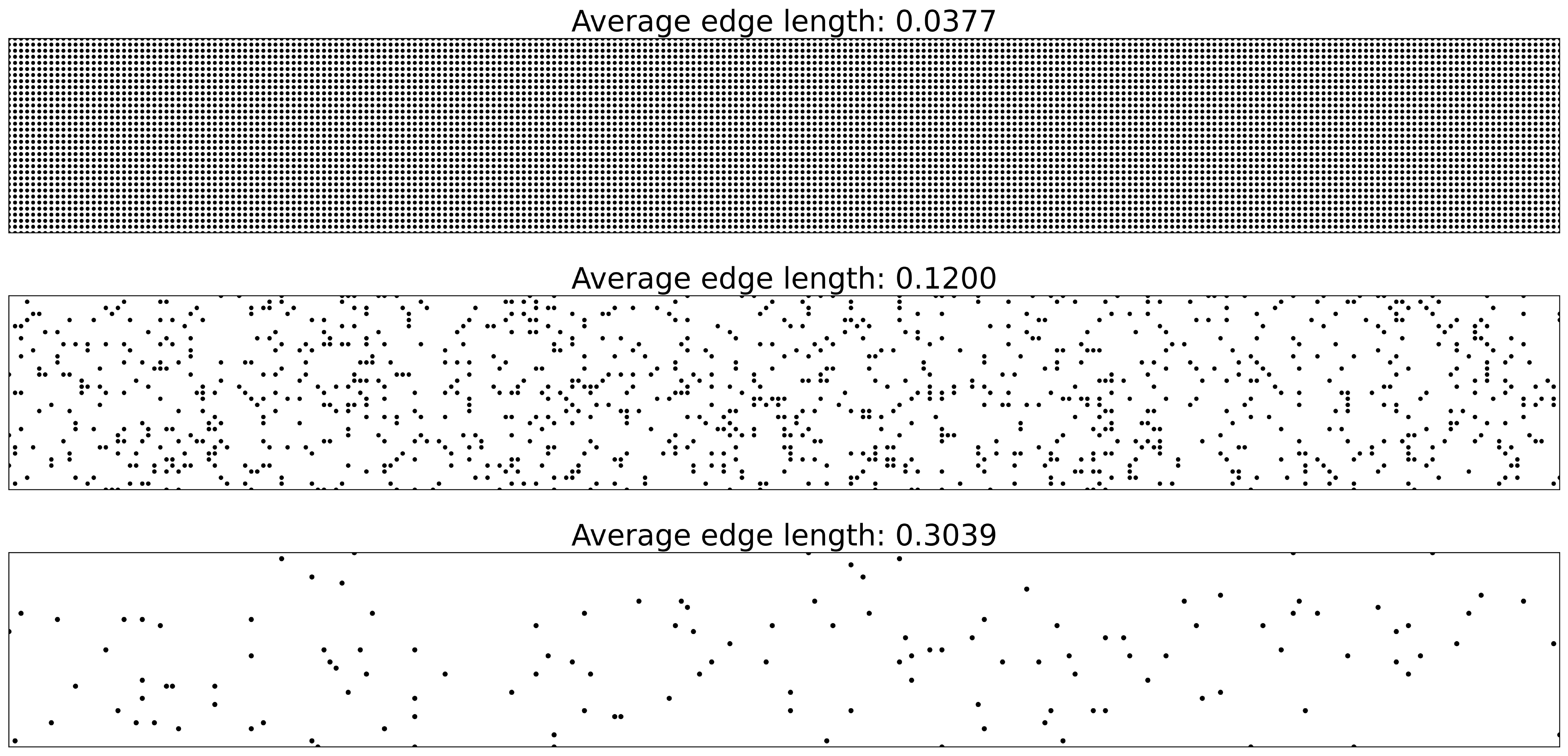}
  \end{center}
\caption{Examples of the multiscale point clouds generated during forecast predictions of Marsigli flow evolution. The finest graph, used in the loss computation, is at the top, and the graphs below are sequentially sampled from this top-level graph. We report the average edge length of the graphs generated, which increases with subsequent sampling.}
  \label{fig:multilevel}
\end{figure}

Fig. \ref{fig:multilevel} shows an example of the multiscale graphs that are generated at each integration step in the prediction. The top level graph is obtained directly from the ground truth data, in this case a uniform grid. During each of the coarsening steps, a random subset of nodes is selected for use in the coarse graph, which is generated using a larger radial cutoff. We can validate that this approach is capturing multiscale features by examining the edge lengths at each graph level. The coarser graphs should have longer edge lengths such that information is distributed farther across the domain during each message-passing layer. We present the average edge length for this set of graphs and indeed see that the average increases with each coarsening step.

\section{Conclusions}

In this work, we develop a rotation-equivariant multiscale graph neural network framework for modeling spatiotemporal fluid dynamics in arbitrary domains. We evaluate two equivariant architectures, one that encodes a latent space composed of scalar and geometric vector features, and one that is considerably computationally cheaper, encoding an invariant latent space of scalar quantities. We introduce spatial multiscale features by incorporating graph coarsening layers, which have been designed to have limited computational burden. The equivariant networks are compared with near-equivalent non-equivariant analogues to understand the role equivariance plays in fluid forecasting. The effect of imbuing equivariance through data augmentation is also contrasted with the hard-constrained equivariant architectures.

Model performance is assessed on two two-dimensional fluid datasets, laminar cylinder flow confined in a channel and the lock-exchange problem following the Boussinesq equations. Both datasets are obtained from numerical simulation of the governing equations, and the point cloud used for modeling is taken from the computational mesh. Models are trained separately on both invariant and non-invariant representations of the flow state to predict the next-step time evolution of the flow field. In the evaluation phase, models are tested by forecasting many future timesteps using an iterative rollout on a test set where either the initial conditions or system parameters are unseen during training.

We find that the use of invariant representations provides a significant benefit to the generalization task. Modeling the vorticity, streamfunction, and pressure or temperature field resulted in higher forecast accuracy in nearly all cases. We also see that data augmentation, while capable of reducing the unsupervised equivariance error by several orders of magnitude, is not particularly effective at generating accurate forecasts. Only in the case where all inputs to the network were scalar did data augmentation lead to a forecast $R^2>0.98$. Furthermore, embedded equivariance can improve forecast accuracy even relative to non-equivariant networks that have been trained and tested on one data orientation. Lastly, we show that the equivariant network with invariant encoder achieves comparable or often better accuracy than the equivariant network with equivariant encoder at a substantially reduced training and evaluation cost.

In summary, this study shows the important role of rotation-equivariance and invariance in spatiotemporal modeling of fluid flows with graph neural networks. 
These models must ultimately learn to recognize the complex relationships within the data, hard-constraining known relationships like equivariance and invariance can reduce the model design space and lead to improved generalization. One major drawback to equivariant GNNs has been their increased computational cost. Here, we show that computationally cheaper scalar latent spaces can also be effective for modeling 2D flows. In fluid forecasting tasks, we suggest modeling invariant quantities, with data-driven invariant encoders serving as a viable alternative if invariant representations are unknown.

\section{Acknowledgements}
This material is based upon work supported by the National Science Foundation Graduate Research Fellowship under Grant No. DGE 1745016 awarded to VS. The authors from CMU acknowledge the support from the Technologies for Safe and Efficient Transportation University Transportation Center, and Mobility21, A United States Department of Transportation National University Transportation Center. This work was supported in part by Oracle Cloud credits and related resources provided by the Oracle for Research program. RM acknowledges support from DOE-SC-FOA-2493 - ``Data-intensive Scientific Machine Learning". This material is based upon work supported by the U.S. Department of Energy (DOE), Office of Science, Office of Advanced
Scientific Computing Research, under Contract DE-AC02-06CH1135.

\clearpage
\bibliography{bib}

\end{document}